\documentclass{aa}  

\usepackage{graphicx}
\usepackage{multirow}
\usepackage{amsmath}
\usepackage{txfonts}
\usepackage{natbib}
\bibpunct{(}{)}{;}{a}{}{,} % to follow the A&A style
\def\ltsim{\raise 2pt \hbox {$<$} \kern-1.1em \lower 4pt \hbox {$\sim$}}
\def\gtsim{\raise 2pt \hbox {$>$} \kern-1.1em \lower 4pt \hbox {$\sim$}}

\begin{document} 

\title{The mystery of the `Kite' radio source in Abell 2626: \\ 
 insights from new Chandra observations}

\author{A. Ignesti\inst{1,2}, M. Gitti\inst{1,2}, G. Brunetti\inst{2}, E. O'Sullivan\inst{3}, C. Sarazin\inst{4}, K. Wong\inst{5,6}}

\institute{
  Dipartimento di Fisica e Astronomia, Universit\`a di Bologna, via Gobetti 93/2, 40129 Bologna, Italy \\
  \email{ alessandro.ignesti2@unibo.it} 
\and 
INAF, Istituto di Radioastronomia di Bologna, via Gobetti 101, 40129 Bologna, Italy 
\and
Harvard-Smithsonian Center for Astrophysics, 60 Garden Street, Cambridge, MA, 02138, USA
\and
Department of Astronomy, University of Virginia, 530 McCormick Road, Charlottesville, VA 22903, USA
\and
%Naval Research Laboratory, 4555 Overlook Avenue SW, Code 7213, Washington, DC 20375, USA
%\and
%Department of Astronomy, University of Maryland, College Park, MD 20742, USA
%\and
Department of Physics and Astronomy, Minnesota State University, Mankato, MN 56001, USA
\and
Eureka Scientific, Inc., 2452 Delmer Street Suite 100, Oakland, CA 94602-3017, USA
%\and
%National Centre for Radio Astrophysics, Tata Institute of Fundamental Research, Post Bag 3, Pune 411007, India
}
\authorrunning{Ignesti et al.}
\titlerunning{The mystery of the `Kite' radio source in Abell 2626: \\ 
 insights from new Chandra observations}

\date{Accepted }

\abstract 
{%Context: 
  We present the results of a new Chandra study of the galaxy cluster Abell~2626. The radio emission of the cluster shows a complex system of four symmetric arcs without known correlations with the thermal X-ray emission. The mirror symmetry of the radio arcs toward the center and the presence of two optical cores in the central galaxy suggested that they may be created by pairs of precessing radio jets powered by dual AGNs inside the core dominant galaxy. However, previous observations failed to observe the second jetted AGN and the spectral trend due to radiative age  along the radio arcs, thus challenging this interpretation.}
{%Aims:
  The new Chandra observation had several scientific objectives, including the search for the second AGN that would support the jet precession model. We focus here on the detailed study of the local properties of the thermal and non-thermal emission in the proximity of the radio arcs, in order to get more insights into their origin.}
{%Methods:
   We performed a standard data reduction of the Chandra dataset deriving the radial profiles of temperature, density, pressure and cooling time of the intra-cluster medium. We further analyzed the 2-D distribution of the gas temperature, discovering that the south-western junction of the radio arcs surrounds the cool core of the cluster. }
{%Results:
   We studied the X-ray surface brightness and spectral profiles across the junction, finding a cold front spatially coincident with the radio arcs. 
  This may suggest a connection between the sloshing of the thermal gas and the nature of the radio filaments, raising new scenarios for their origin. A tantalizing possibility is that the radio arcs trace the projection of a complex surface connecting the sites where electrons are most efficiently reaccelerated by the turbulence that is generated by the gas sloshing. In this case, diffuse emission embedded by the arcs and with extremely steep spectrum should be most visible at very low radio frequencies.} {}


\maketitle

\vspace{-0.15in}
\section{Introduction} 
%\vspace{-0.05in}

Chandra high-resolution X-ray images of many merging and relaxed clusters revealed that the sloshing of cold sub-clumps of the intra-cluster medium (ICM) may form prominent contact discontinuities, or “cold fronts” \citep[e.g.,][]{Markevitch-Vikhlinin_2007}. Cold fronts are seen as sharp edges in the cluster X-ray images and temperature maps, and they are identified as temperature discontinuities in pressure balance with the ICM. On the other hand, radio observations revealed that a number of cool-core clusters host a radio mini-halo -- a faint, diffuse radio source with a steep spectrum ($\alpha < -1$, where the flux density as a function of frequency is $S_{\nu} \propto \nu^{\alpha}$) and a size comparable to that of the cool-core region \citep[e.g.,][]{Feretti_2012, Giacintucci_2017}. A number of mechanisms have been proposed for the origin of the relativistic, radio-emitting electrons \citep[e.g.,][ for leptonic and hadronic models, respectively]{Gitti_2002, Pfrommer-Ensslin_2004}, but the physics of the mechanisms that produces mini-halos is still debated \citep[e.g.,][]{Brunetti-Jones_2014}.

The galaxy cluster Abell 2626 (hereafter A2626) is a low-redshift \citep[z=0.0553][]{Struble-Rood_1999}, regular, poor cluster, with
%It is located at RA 23h36m30s, DEC+21d08m33s and it is part of the Perseus-Pegasus super-cluster. A2626 has
an estimated mass of 1.3$\times10^{15}$ M$_{\odot}$ and a virial radius of 1.6 Mpc \citep[][]{Mohr_1996}. The most luminous galaxy is the central dominant (cD) galaxy IC5338, which hosts a pair of optical nuclei with a projected separation of $\sim$3.5 arcsec, of which only the southern one has a counterpart also in the radio \citep[][]{Owen_1995} and hard X-ray bands \citep[see][Fig. 3]{Wong_2008}. A2626 also hosts the S0 galaxy IC5337, that has been previously classified as a jelly-fish galaxy by \citet[][]{Poggianti_2016} for its optical properties. In the X-ray band, A2626 appears as a relaxed cluster with a roundish morphology. It is a cool-core cluster with estimated X-ray luminosity of $1.9 \times 10^{44}$ erg s$^{-1}$ and mass accretion rate of 4 M$_{\odot}$ yr$^{-1}$ \citep[e.g.,][]{Bravi_2016}. On the other hand, in the radio band A2626 shows a peculiar, extended emission with arc-like, symmetric features forming a striking kite-like global morphology \citep[][]{Gitti_2004, Gitti_2013b,Kale_2017, Ignesti_2017}, whose origin is still unclear. Initially, \citet[][]{Gitti_2004} proposed that the cluster hosts a candidate radio mini-halo embedding the northern (N) and southern (S) arcs, whereas \citet[][]{Wong_2008} argued that the such elongated radio features may be produced by jet precession triggered by the reciprocal gravitational interactions of the two cores of the cD galaxy. Their scenario was proposed when only the N ans S arcs were known. However, the subsequent discovery of the third \citep[][]{Gitti_2013b} and fourth \citep[][]{Kale_2017} radio arc to the west (W) and east (E) directions, respectively, and the estimated time scales of the precession period \citep[][]{Ignesti_2017} challenged this model.

In this work we present the analysis of the new Chandra observations of A2626, searching for the evidence of a second AGN that may explain 
%the presence of 
the second pair of radio arcs, and for correlations between the thermal and non-thermal plasma in the radio arc regions.  We adopt a $\mathrm{\Lambda CDM}$ cosmology with $\mathrm{H_{0}=70}$ km $\mathrm{s^{-1}Mpc^{-1}}$, $\Omega_{M} = 1 - \Omega_{\Lambda} = 0.3$. The cluster luminosity distance is 246.8 Mpc, and the angular scale is 1 arcsec = 1.1 kpc\footnote{{\ttfamily
    http://www.astro.ucla.edu/$\#$7Ewright/CosmoCalc.html}}. The confidence level of the reported values is 90$\%$  (1.64 $\sigma$).

\vspace{-0.15in}
\section{Observation and Data Reduction}
%\vspace{-0.05in}

A2626 has been observed with the Chandra space observatory in October 2013 for 110 ks (ObsID: 16136, PI: C. Sarazin). The dataset was reprocessed with CIAO v4.8 using CALDB v2.7.2 and corrected for known time-dependent gain and charge transfer inefficiency problems following techniques similar to those described in the Chandra analysis threads\footnote{{\ttfamily http://cxc.harvard.edu/ciao/threads/index.html}}. In order to filter out strong background flares, we also applied screening of the event files, slightly reducing the effective observation time to 109 ks. We used blank-sky background files normalized to the count rate of the source image in the 10-12 keV band for background subtraction. Finally, we identified and removed the point sources using the CIAO task {\ttfamily WAVDETECT}, with the detection threshold set to the default value of $10^{-6}$ .

\vspace{-0.15in}
\section{Results}

\begin{figure*}
 \includegraphics[width=1\textwidth]{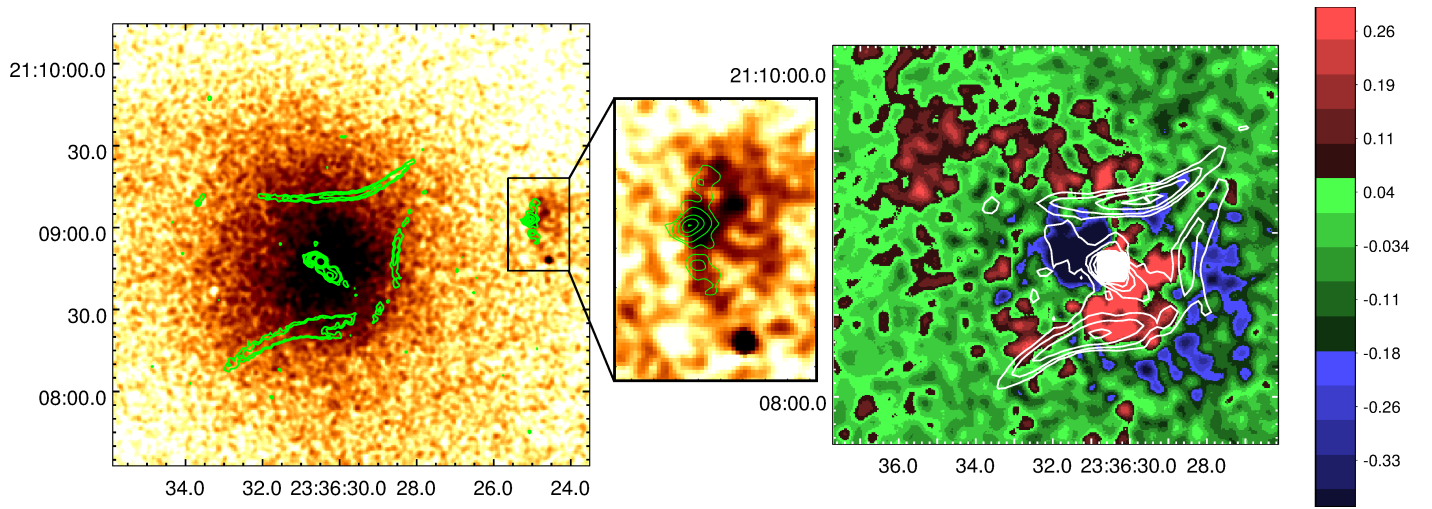}
 \caption{\label{x1.fig}{\it Left:} 0.5-2 keV Chandra image of A2626 smoothed with a 3 pixel gaussian filter (1 ACIS pixel = 0.5$''$), with a zoom on IC5337 and with the 1.4 GHz radio emission overlaid in green contours \citep[from Fig. 2 of][resolution $\sim$1.2$''$]{Gitti_2013b}. {\it Right:} SB residual map obtained by subtracting the $\beta$-model, smoothed with a 8 pixel gaussian filter (color map), and with the 1.4 GHz radio emission overlaid in white contours \citep[from Fig. 3 of][resolution $\sim$4.2$''$]{Gitti_2013b}. As indicated by the color bar on the right, the over-densities, in unit of counts px$^{-2}$s$^{-1}$, are shown in red and the sub-densities in blue.}
\end{figure*}

In this work we focus on the results of our analysis concerning the correlations of the local properties of the ICM with the radio arcs, in order to get more insights into their origin. In particular, after an analysis of the global properties of the cluster, we analyzed the 2-D distribution of the thermal properties of the ICM in the proximity of the radio arcs. In Figs. \ref{x1.fig} and \ref{ewan_T.fig} we overlay the 1.4 GHz radio emission (from Gitti 2013) imaged at slightly different resolutions, depending on the specific aim of each image.

\vspace{-0.15in}
\subsection{Global X-ray properties}
\vspace{-0.05in}
We report in Fig. \ref{x1.fig} (left panel) the 0.5-2.0 keV raw image of A2626 smoothed with a gaussian filter. We detected three point sources associated with A2626, the southern core of IC5338 (23h 36m 30.5s, +21d 08m 47.7s), the core of IC5337 (23h 36m 25.0s, +21d 09m 2.9s) and one in the south tail of IC5337 (23h 36m 24.6s, +21d 08m 48.4s) that does not have a radio counterpart. We did not detect any hard (2.0-10.0 keV) X-ray emission associated to northern core of IC5338, in agreement with the results of \citet[][]{Wong_2008}. By considering the contribution of the thermal emission of the ICM and the instrumental background, we could estimate an upper limit for its X-ray luminosity of $1.3\times10^{40}$ erg s$^{-1}$ (0.5-7.0 keV band). Therefore, we cannot find evidence that the second pair of radio arcs are due to a second AGN, although it could be that the AGN has just turned off.

\begin{figure}
 \includegraphics[width=.5\textwidth]{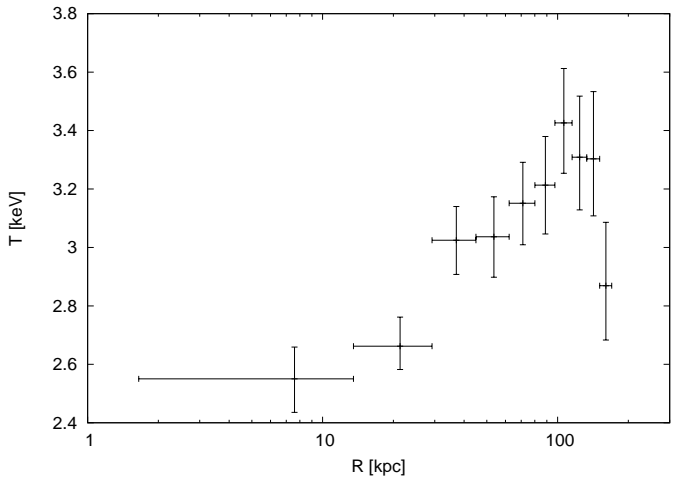}
 \caption{\label{temp.fig}  Azimuthally-averaged projected radial temperature profile of A2626 inside 170 kpc. }
\end{figure}
We extracted the azimuthally-averaged radial temperature profile in circular annuli (Fig. \ref{temp.fig}), that reveals that the temperature decreases from a maximum value of $\sim$3.4 keV to $\sim$2.6 keV towards the center, with a discontinuity at $\sim$30 kpc. We observed a second temperature drop in the outermost bin at $\sim$170 kpc. The possible cause of this drop will be discussed in Section 3.2 .

In order to study the dynamical state of the cluster, we estimated the cooling time radial profile $t_{c}(r)$ based on the temperature and density radial profiles. The cooling time $t_{c}$ is the time-scale of the cooling processes acting in the ICM and it can be derived as follows \citep[e.g.,][]{Gitti_2012}:
\begin{equation}
 t_{c}(r)=\frac{\gamma}{\gamma-1}\frac{kT}{\mu X n_{e}(r)\Lambda(T(r))}
 \label{tcool.math}
\end{equation}
where $\gamma=5/3$ is the adiabatic index, $\mu\simeq0.6$ is the mean molecular weigh, $X\simeq0.7$ is the hydrogen mass fraction of the ICM, $n_{e}(r)$ and $\Lambda(T(r))$ are the electron density of the ICM and the cooling function \citep[e.g.,][]{Sutherland-Dopita_1993} at the radius $r$. 
The radius at which the cooling time is 3 Gyr is typically called cooling radius, $R_{C}$, and it delimits the cool core that is the region where the cooling processes are efficient. We estimated $R_{C}\simeq$23 kpc, in agreement with \citet[][]{Bravi_2016}. 
We then extracted the spectrum of the whole cluster inside $R_{C}$ to estimate its cooling luminosity and the mass accretion rate. We fitted the spectrum with the {\ttfamily wabs$\cdot$(apec+mkcflow)} model in Xspec v.12.9, where the {\ttfamily mkcflow} model takes into account an isobaric multi-phase cooling flow component in addition to the thermal emission of the ICM modeled by {\ttfamily apec} \citep[][]{Arnaud_1996}. 
%
% In order to consider the standard cooling flow model, the higher temperature
%component ($kT_{high}$) of the {\ttfamily MKcold frontLOW} model was tied to the thermal model,
%under the assumptions that the thermal component represents the ambient
%cluster gas and that cooling flow gas is cooled ambient gas. The lower temperature component ($kT_{low} $) was fixed close to zero, namely at 0.0808 keV.\\ 
%
We concluded that A2626 is a weak cooling flow cluster, with a mass accretion rate of 2.5$\pm$0.8 M$_{\odot}$ yr$^{-1}$ and a cooling luminosity of 4.2$\times 10^{43}$ erg s$^{-1}$ in the 0.5-7 keV energy band. The mean temperature inside $R_{C}$ is 2.4$\pm$0.1 keV. These results are in agreement with \citet[][]{Wong_2008} and \citet[][]{Bravi_2016}.

 In order to search for correlations with the radio arcs, we carried out a morphological analysis by subtracting a 2-D $\beta$-model \citep[][]{Cavaliere-Fusco_1976} of surface brightness (SB) distribution to highlight over-densities or depressions in the ICM distribution. We performed this operation with the software SHERPA of CIAO v4.9. The best-fit values for the core radius, $R$, the index $\beta$, and the amplitude, $A$, are $R=26.1\pm0.5$ arcsec, $\beta=0.400\pm0.003$, and $A=4.86\pm0.08$ counts s$^{-1}$ arcmin$^{-2}$. The SB model has been centered on the southern nucleus of IC5338 and the emission of the core of IC5338 and IC5337 has been masked because it was not related to the cluster thermal emission. We also performed a fit with a double $\beta$-model, usually adopted for cool-core clusters \citep[e.g.,][]{Xue_2000}, finding that the addition of a second component does not significantly improve the fit.

In Fig.\ref{x1.fig} (right panel) we report the residual map with the 1.4 GHz radio contours overlaid \citep[from Fig. 3 of][]{Gitti_2013b}. This map shows that there is an emission excess, which is related to an over-density of the ICM, that is remarkably delimited by the south-western (SW) junction of the radio arcs.  On a larger scale to the north, the map shows another over-density with a spiral-like morphology. We also observe a depression on the east of the core that was already noticed by \citet[][]{Shin_2016}, who argued that it may be related to AGN activities. We are cautious about claiming the detection of a cavity, because we cannot exclude that it may be an artifact produced by the asymmetrical SB distribution around the core, similar to the sub-density seen outside of the SW junction. However, the spatial connection of the X-ray depression with the central radio emission (Fig. \ref{x1.fig}, right panel) may support its interpretation as a radio-filled cavity.

%\begin{figure}
% \includegraphics[width=.5\textwidth]{ic5337X.eps}
%\end{figure}
%\subsection{Analysis of the cores of IC5338}
%\begin{figure}
% \includegraphics[width=.5\textwidth]{2.eps}
%\end{figure}

\vspace{-0.15in}
\subsection{Spectral temperature map}
\vspace{-0.05in}

\begin{figure*}
\centering
 \begin{minipage}[t]{1\textwidth}
  \centering
   \includegraphics[width=1\textwidth]{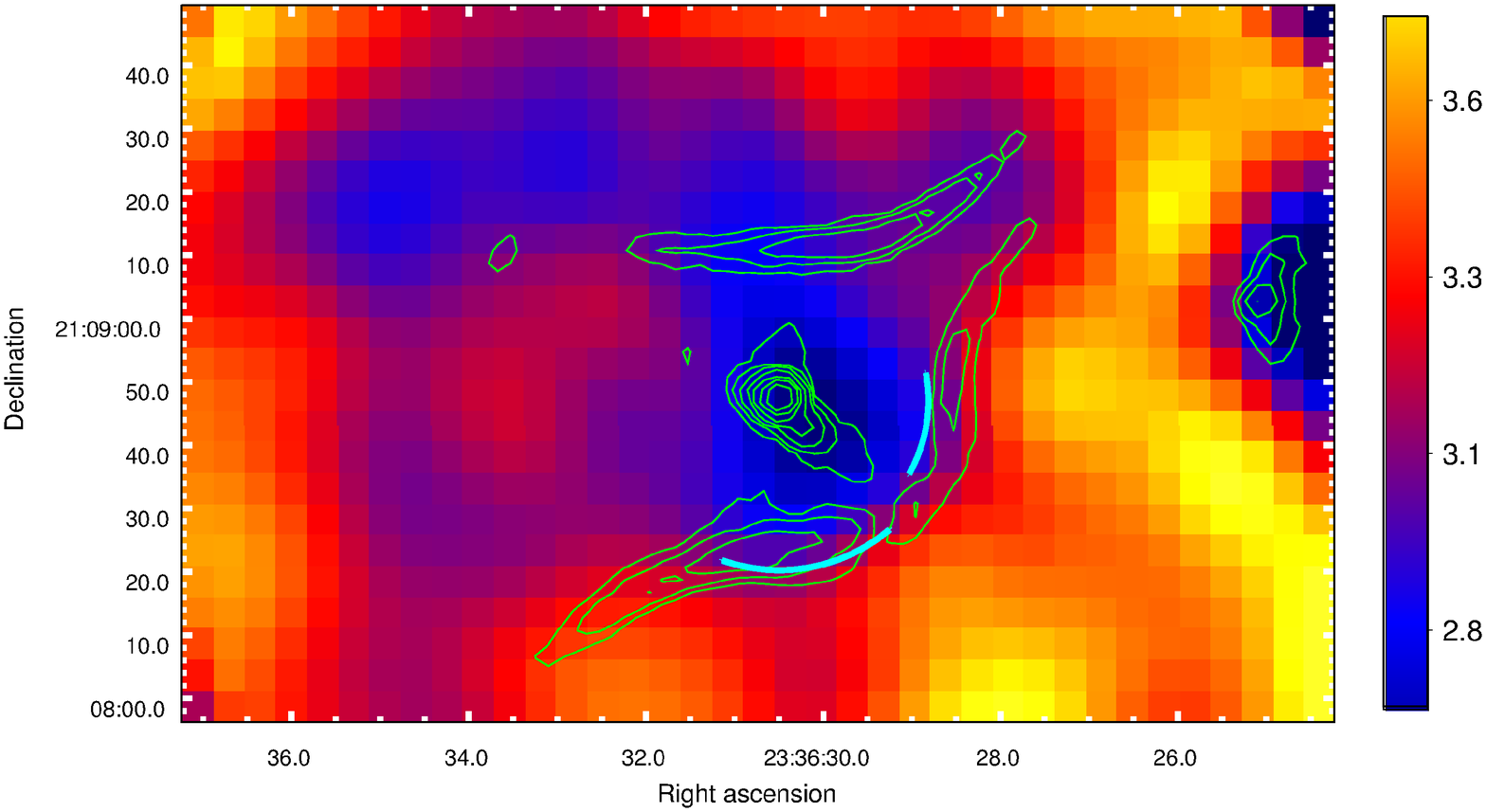}
   \end{minipage}%
   \hspace{5cm}
   \vspace{0.5cm}
 \begin{minipage}[r]{.5\textwidth}
  \centering
   \includegraphics[width=1\textwidth]{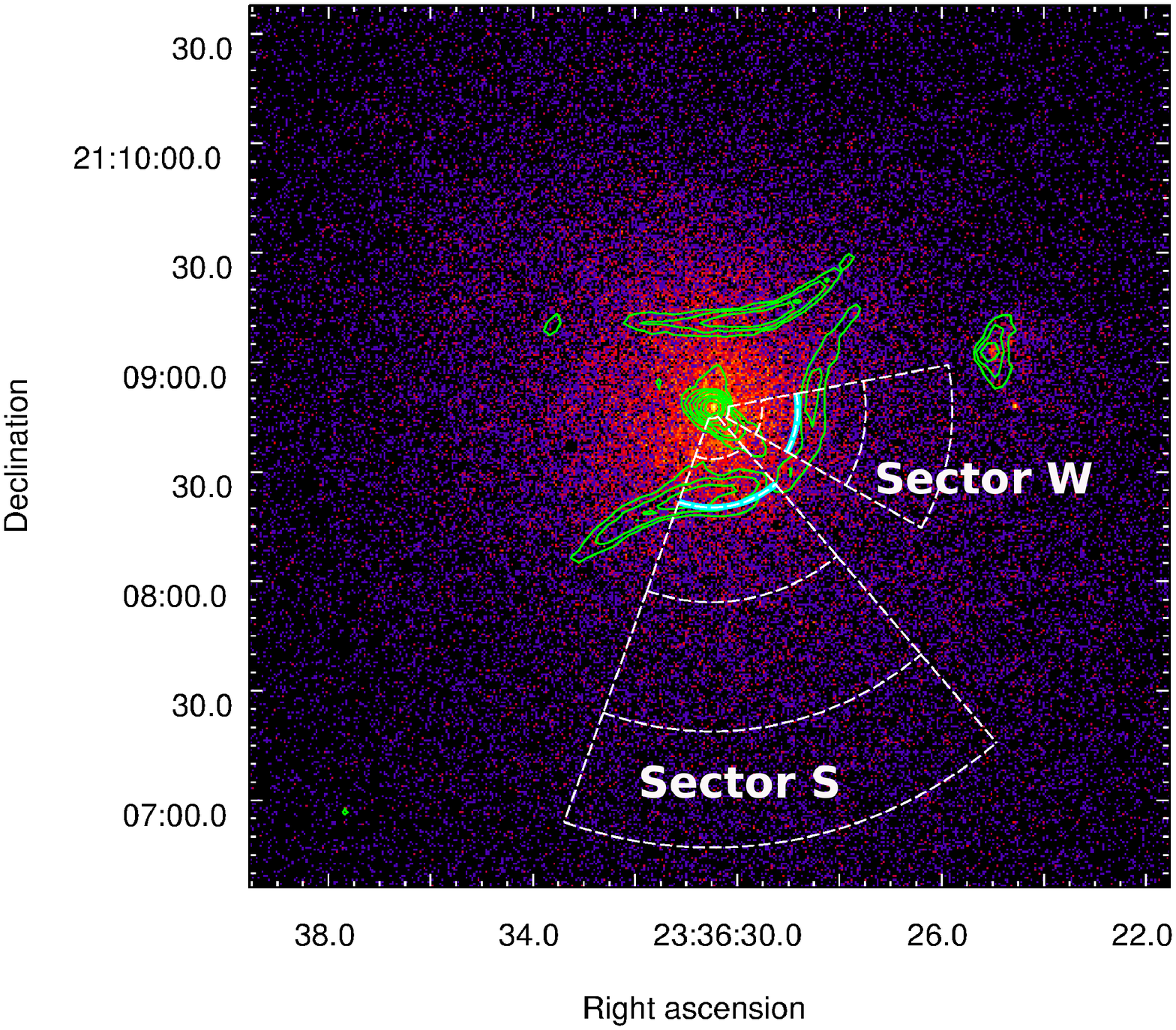}
   \end{minipage}%
   \hspace{0.2cm}
   \begin{minipage}[l]{.48\textwidth}
  \centering
   \includegraphics[width=1\textwidth]{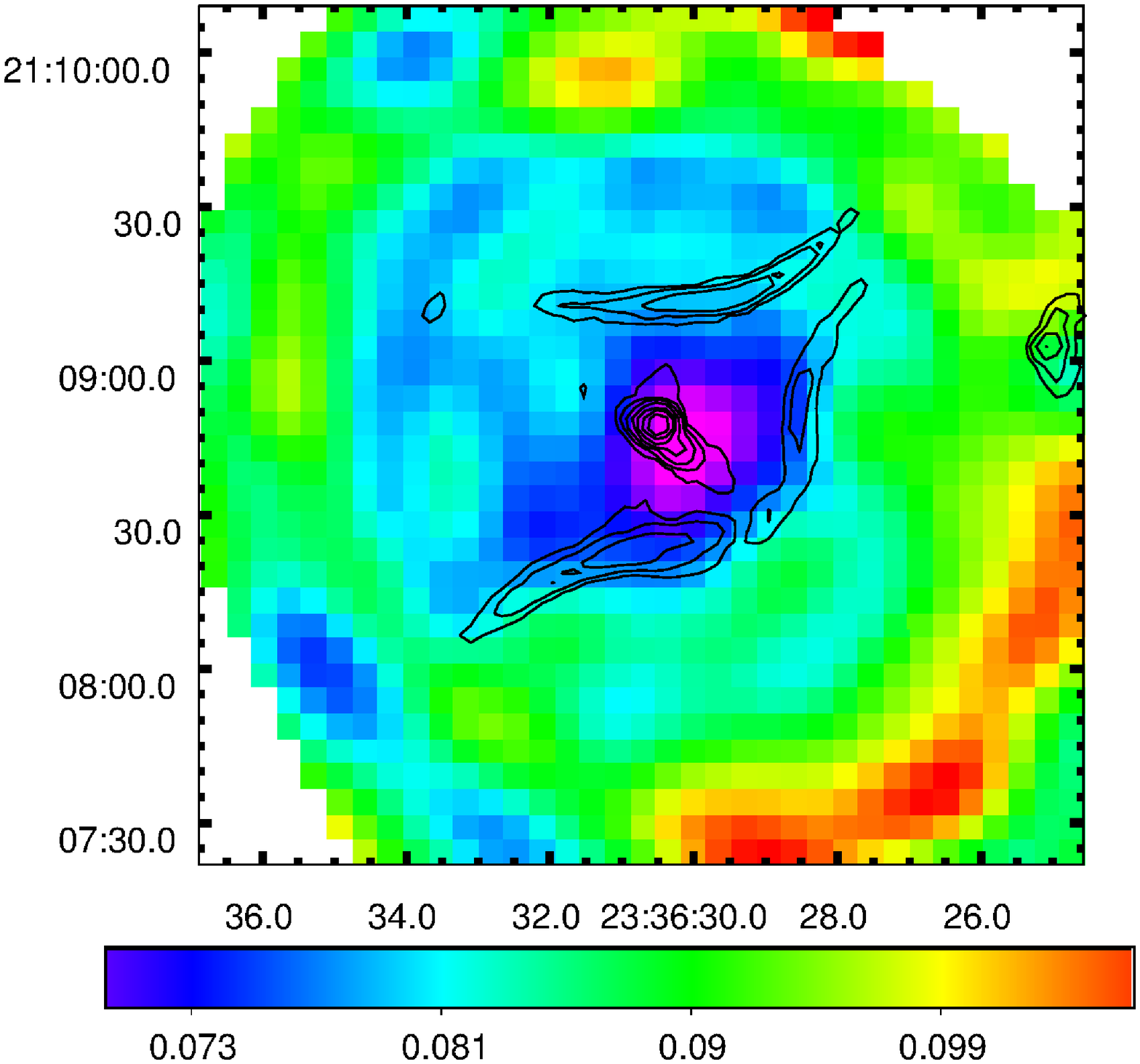}
   \end{minipage}%
   \caption{\label{ewan_T.fig}  {\it Top:} Temperature map of A2626. The color bar indicates the ICM temperature in keV. In green are reported the contours of the 1.4 GHz radio emission from \citet[][resolution $\sim$3.3$''$]{Gitti_2013b}, in cyan the position of the cold fronts determined by our analysis. {\it Bottom-left:}Raw image of A2626 . In green are reported the same 1.4 GHz contours as in the top panel, in white the sectors where we extracted the spectral profiles, where we have highlighted in cyan the position of the SB jumps. {\it Bottom-right:} Relative temperature error map. In black are reported the same 1.4 GHz contours as in the top panel.}
\end{figure*}

The spectral temperature map shown in Fig. \ref{ewan_T.fig} (top) was created by using the techniques 
described in \citet[][]{O'Sullivan_2011a}. The map pixels are $\sim 5''$ (10 ACIS physical pixels) square. The value in each pixel is the best-fitting temperature obtained from an absorbed {\ttfamily apec} model fit to a spectrum extracted from a circular region centred on the pixel, with a radius chosen to ensure that it contains at least 1500 net counts. The effective resolution of the map is therefore determined by the size of the extraction regions, whose radii range from $\sim 5''$ in the cluster centre to $\sim20''$ in the outskirts. Since the spectral extraction regions
are larger than the map pixels, individual pixel values
are not independent and the maps are, thus, analogous to adaptively smoothed images, with more smoothing in regions of  lower
surface brightness. In the bottom-right panel of Fig. \ref{ewan_T.fig} we report the relative error map. To produce this map we divided the mean error in each pixel, which corresponds to the arithmetic mean between the 1-$\sigma$ upper and lower bounds, by the best-fit temperature in that pixel. Uncertainties on the kT measurements range from $\sim7\%$ to $\sim10\%$.

%Due to above assumptions, the values of the these maps 
%are approximate, and they should be supported with the spectral profiles. Nevertheless they can hint to the presence of substructures, that may be not 
%detected in a surface brightness map. \\
In Fig. \ref{ewan_T.fig} (top panel) we also report the contours of the radio emission at 1.4 GHz to identify possible correlations between the radio arcs and the ICM. This map shows that the ICM exhibits an inner cold region with $T_{cold}<$2.8 keV and an outer hot region with $T_{hot}\sim$3.4 keV. Remarkably, the edge of these regions coincides with the SW junction of the radio arcs. On the opposite direction, the cold ICM is not confined to the inner part of A2626, but it is elongated outside in a spiral-like shape that coincides with the northern spiral feature observed in the SB residual map (Fig. \ref{x1.fig}, right panel) and resembles the cold spirals often related to the sloshing of the ICM \citep[e.g.][]{Markevitch-Vikhlinin_2007}.

The temperature map also shows that IC5337 is leaving a trail of cold plasma ($T\sim$2.3 keV). Our result, in agreement with \citet[][]{Wong_2008}, may confirm that the jelly-fish galaxy is losing its cold atmosphere due to the interactions with the ICM. The accurate study of IC5337 is the object of a paper in preparation.

Lastly, we note the presence of a cold feature to the south-west edge of the map (not shown in Fig. \ref{ewan_T.fig}, top panel) located at the same radial distance as the last annulus of the temperature profile (Fig. \ref{temp.fig}). In order to estimate the contribution of this feature to the mean temperature of the last temperature bin, we divided the last annulus in different sectors, including and excluding this feature, and compared the resulting spectral fits. In particular, in a 270 degree-wide sector excluding the feature we measured a temperature of $3.4\pm0.2$ keV, which is higher than the value that we report in the azimuthally-averaged radial profile ($2.9\pm0.2$ keV). Therefore, we may conclude that the presence of this feature is likely to be the cause of the temperature drop observed in the outermost bin at $\sim$170 kpc.

%\subsection{Surface brightness residual map}
%\begin{figure}
% \includegraphics[width=.5\textwidth]{re.eps}
% \end{figure}
%
%We further performed a 2D morphological analysis of the cluster emission by fitting the SB of the whole cluster with 
%a double $\beta$-model. If the process that originated the arcs interacted also with the ICM, then the plasma may "remember" that event by exhibiting an anomalous distribution of the SB. We carried out this analysis on a circular region centered on IC5337, where we removed the emission of IC5337 and the core of IC5338. We report the residual map
%and 
%the best-fit parameters in Fig. \ref{residui.fig}.\\ 
%The 2D fit was performed with the software SHERPA v. 4.4 of the CHANDRA data analysis package CIAO. The software fits the brightness value of every pixel of a map with a model. It provides the best fit parameters and the residual map, obtained by subtracting the model from the image. This map highlights the regions whose emission differs from the model, pointing out every over-densities, or sub-densities, in surface brightness.
%The residual map (Fig. \ref{residui.fig}) hints at the presence of a cavity in the emission to the east, and an excess to the west. It is interesting to notice 
%that, again, the SW junction of the radio arcs delimits the emission excess.\\ 
%The temperature jump (Fig. \ref{ewan_T.fig}) and the SB excess (Fig. \ref{residui.fig}) suggest the presence of possible edges in the ICM. 

\subsection{Analysis of the south-west junction}

\begin{figure*}
\centering
 \begin{minipage}[l]{.48\textwidth}
  \centering
   \includegraphics[width=1\textwidth]{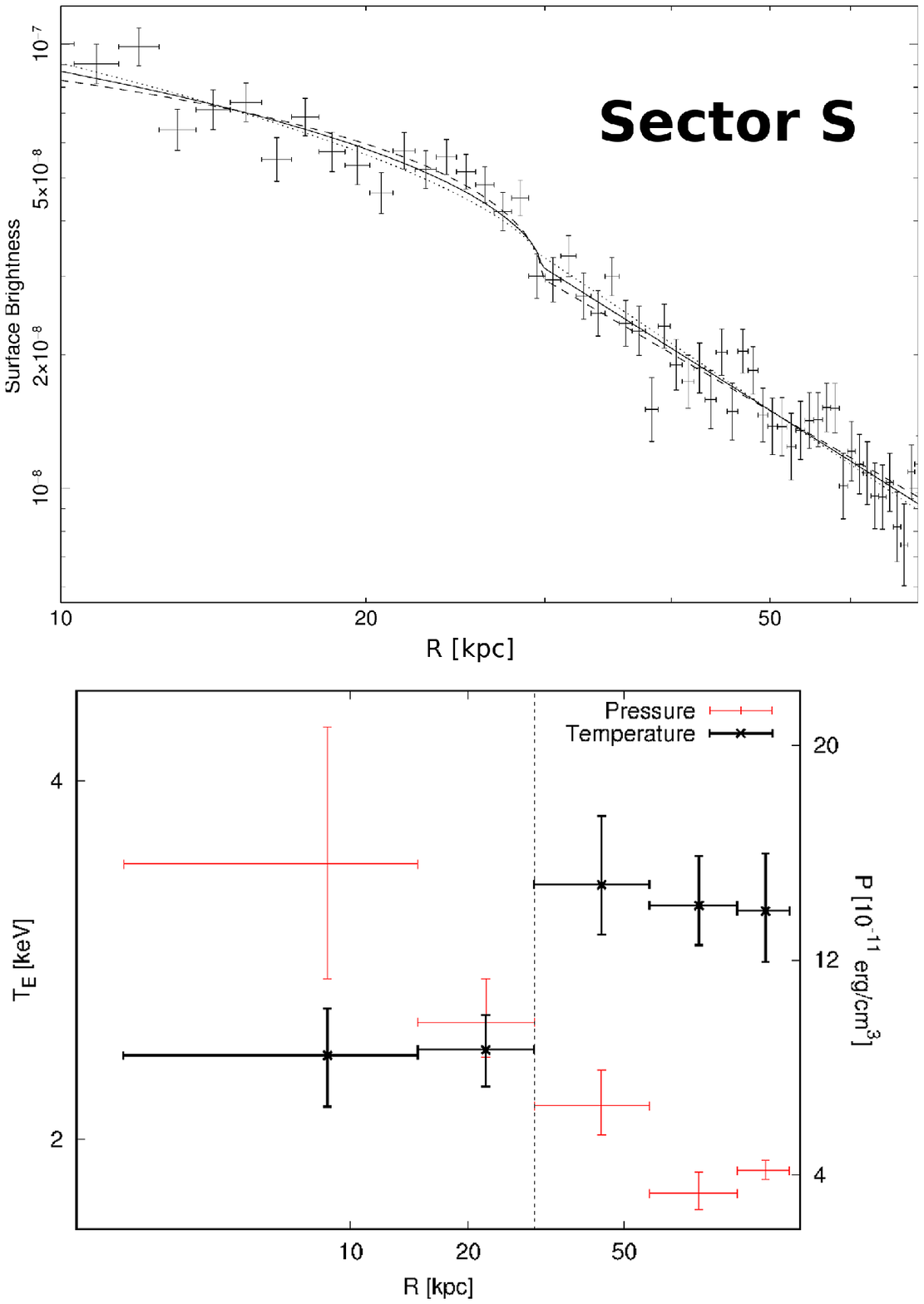}
   \end{minipage}%
   \hspace{0.5cm}
 \centering
 \begin{minipage}[r]{.48\textwidth}
  \centering
   \includegraphics[width=1\textwidth]{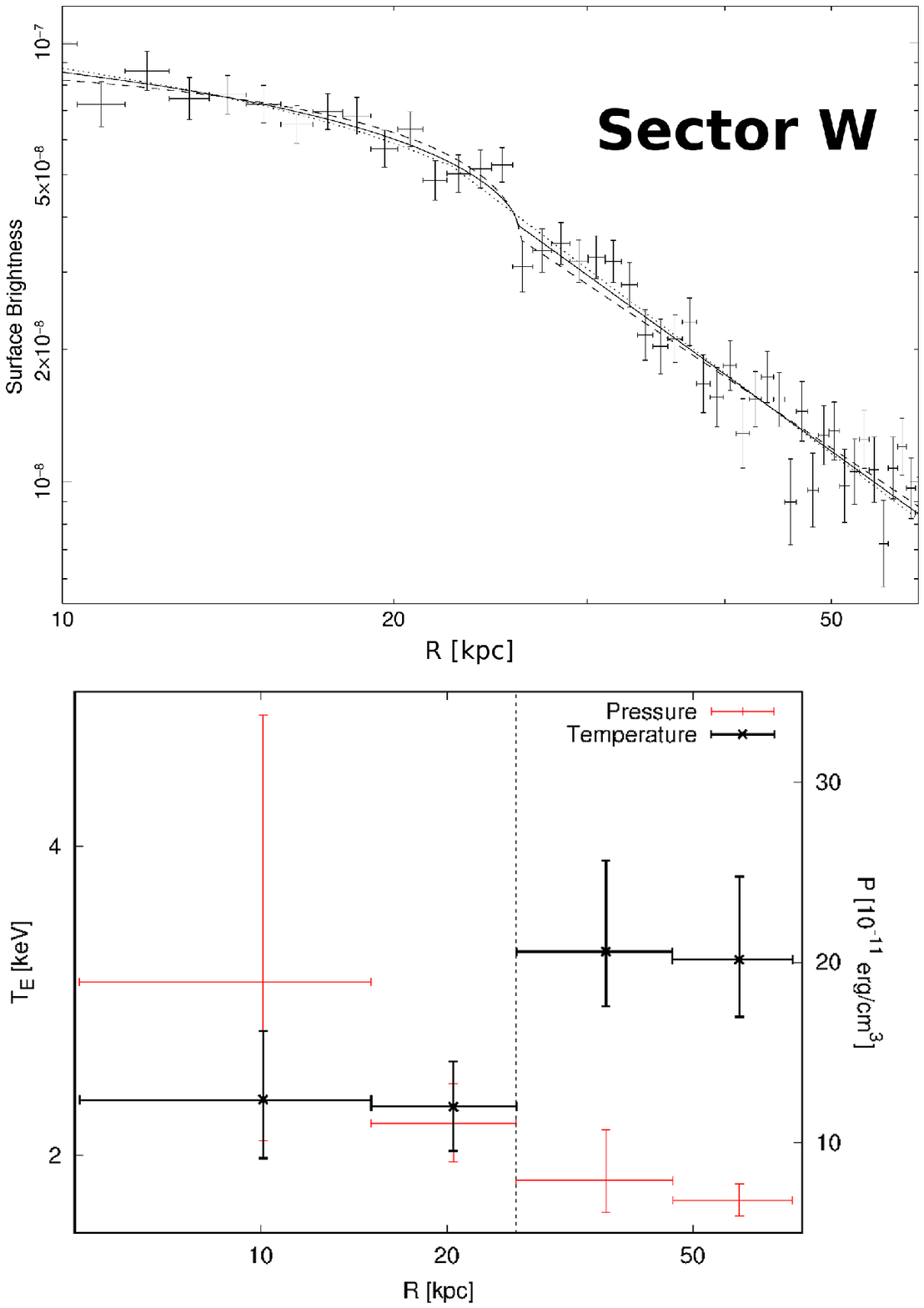}
   \end{minipage}%
   \caption{\label{prof-sw.fig}Radial profiles observed in the southern (left) and western (right) sectors reported in Fig. \ref{ewan_T.fig}. In each panel: {\it Top:} SB profile with the best-fit to the broken power-law model. The SB is expressed in unit of counts px$^{-2}$ s$^{-1}$, the radius is in units of kpc. The dashed lines show the upper and lower confidence bounds of the best-fit density ratios (corresponding to 1.32$_{-0.13}^{+0.15}$ for the Sector S and to 1.37$_{-0.17}^{+0.19}$ for the Sector W). {\it Bottom:} Projected temperature (black) and de-projected pressure (red) profiles.% On the left y-axis is shown the temperature scale and on t he right y-axis the pressure scale.
   The black dashed line indicates the position of the front determined by the broken power-law fit to the SB profile shown in the upper panels.}
\end{figure*}

The SB residual (Fig. \ref{x1.fig}, right panel) and the spectral temperature (Fig. \ref{ewan_T.fig}, top) maps show that the ICM physical properties inside the SW radio arcs are different than outside, so we focused our analysis on the region across the SW junction to understand why.

By following the geometry of the cold region observed in Fig. \ref{ewan_T.fig}, we divided the SW cluster region into several sectors. In these regions we extracted the background-subtracted, exposure-corrected SB radial profiles and we fitted them with a broken power-law model \citep[as described in][]{Nulsen_2005a}, in order to determine the position of the SB jumps that we observe in the 2-D residual map (Fig. \ref{x1.fig}, right panel).  In two sectors to the south and to the west, which are showed in the bottom-left panel of Fig. \ref{ewan_T.fig} labeled as `Sector S' and ` Sector W', we detected a SB jump located at a distance from the core of 29.5 kpc and 26.4 kpc, respectively.  They are associated to a ICM density jump of 1.32$_{-0.13}^{+0.15}$ (Sector S) and 1.37$_{-0.17}^{+0.19}$ (Sector W), both detected above the 3$\sigma$ level.  We checked the solidity of our result by varying the extraction sectors (in both angular width and radial binning), always finding a density jump detection above 3$\sigma$.

 We then divided the sectors into several annuli, which are reported in Fig. \ref{ewan_T.fig} (bottom-left panel), to extract the ICM temperature, density and pressure radial profiles, by fitting the spectra of each annulus with both the projected {\ttfamily wabs$\cdot$apec} and the de-projected {\ttfamily projct$\cdot$wabs$\cdot$apec} Xspec models, and paying attention to place the second and the third annuli across the SB jump. We report the temperature and pressure profiles in Fig. \ref{prof-sw.fig}. The profiles show that the SB jumps coincide with a significant drop in the temperature toward the center, from 3.42$_{-0.28}^{+0.38}$ keV to 2.50$_{-0.21}^{+0.19}$ keV in the Sector S and from 3.32$_{-0.36}^{+0.59}$ keV to 2.31$_{-0.29}^{+0.29}$ keV in the Sector W. We observed the temperature drops in both the projected and de-projected profiles with a confidence level above $3\sigma$. The pressure is instead continuous, thus indicating rough pressure balance across the front, that is a signature of the presence of a cold front. Therefore, we argue that A2626 shows a weak cold front in the proximity of the radio arcs. We note that a similar feature has recently been observed in the Perseus cluster \citep[][]{Walker_2017}.

\section{Discussion and future prospects}
 In this work, we have analyzed a new Chandra observation of A2626 in order to get insights on the origin of the radio arcs. We have searched for evidence of the AGN activity of the northern core of IC5338, which has been invoked as a possible origin of the arcs west and east, and we have been able to provide an upper limit for its X-ray luminosity, $L_{X}<1.3\times10^{40}$ erg s$^{-1}$. Moreover, we have analyzed the 2-D distribution of the thermal properties of the ICM in order to find possible correlations between the thermal and non-thermal plasma. In this section, we summarize our work and we propose a new qualitative scenario that may explain the origin of the radio arcs.

The new Chandra data presented here led to the discovery of a cold front in the proximity of the SW radio arcs. This is the most important result in our paper. Indeed the spatial coincidence between the cold front (at least the part of the front that is visible) and the SW radio arc may suggest a connection between the radio arcs and the dynamics of the thermal gas.

One possibility is that the arcs are ghost relativistic bubbles shaped and gently compressed (i.e., at low compression rate) by the motion of the gas.
The observed morphology may thus arise from the action of complex gas motions at the boundary of the sloshing region on a buoyant cloud of relativistic plasma. If on the one hand this might appear a possible explanation for the SW junction, on the other hand the shape of the arc N \citep[and of the arc E seen in][]{Kale_2017} appears  more tricky and would require a customized geometry. It is worth to mention that indeed radio filaments with very steep spectrum are seen in LOFAR observations of galaxy clusters with complex gas dynamics \citep[e.g.][]{Shimwell_2016,de'Gasperin_2017}. The large occurrence of these filaments at low radio frequencies naturally results from their steep spectrum, on the other hand the steep spectrum makes  them very rare at higher frequencies. A2626 might indeed be one of these rare cases, where the seed relativistic bubbles that are compressed and advected by gas motion may originate from the radio activity of the AGNs in the core in the last Gyr of so. 

At the same time, however, we may also speculate that the arcs are the brightest parts of a mini-halo that is generated in the region of the core bounded by the sloshing cold fronts. A connection between mini-halos and cold fronts has been observed by \citet[][]{Mazzotta-Giacintucci_2008}. Numerical simulations show that the sloshing of the gas in these regions amplify the magnetic field within the cold front and generates turbulence that may reaccelerate relativistic particles producing mini-halos \citep[][]{ZuHone_2011,ZuHone_2013}. In this scenario the acceleration rates and magnetic fields vary within the volume of the cluster core. The strongest acceleration rates are observed in many cases near the cold front surfaces of the simulated clusters. In fact simulated images show that mini-halos become increasingly patchy and filamentary at frequencies higher than an observed  synchrotron frequency $\nu_{0}$ \citep[Fig 10-12 in][]{ZuHone_2013}. The maximum steepening frequency $\nu_{s}$ can be obtained  by combining  the equations for the momentum-diffusion coefficients \citep[Eq. 33-34 in][]{ZuHone_2013} with the equation  for the reacceleration time that is requested to generate synchrotron emission at a given frequency in these models \citep[Eq.14 in][]{Brunetti-Lazarian_2016}.  In the case of A2626 we estimated:
\begin{equation}
 \frac{\nu_{s}}{\text{MHz}}\simeq700\left( \frac{v_{t}}{200\text{ km s}^{-1}}\right)^{4}\left( \frac{R^{c}}{0.25}\right)^{2}\left(\frac{<k>}{3.6 \text{ kpc}^{-1}}+\frac{k_{mfp}}{6.3\text{ kpc}^{-1}}\right)^{2}/4 
 \label{nu.math}
\end{equation}
The parameters $v_{t}$, $R^{c}$, $<k>$ and $k_{mfp}$ are the turbulent velocity, the fraction of turbulent energy in the form of compressive modes, the average wavenumber and the wavenumber associated to the mean free path of the particles, respectively \citep[see ][for futher details]{ZuHone_2013}. In Eq. \ref{nu.math} we assumed an homogeneous magnetic field $B=B_{CMB}/\sqrt{3}$. According to Eq. \ref{nu.math}, the frequency of maximum synchrotron emission that we obtain by assuming the normalized values of the parameters  is of the order of 700 MHz. Therefore, an observation at this frequency probing for the presence of the mini-halo will be a crucial test for the turbulent acceleration scenario. 

It may be that in A2626 the sloshing sustains a process of acceleration of electrons that is less efficient than in other mini-halos. Our observations may be picking up only the boundaries of the mini-halo where turbulence is higher and electrons are reaccelerated with an efficiency that is high enough to produce GHz emission. More specifically, according to Eq. \ref{nu.math}, a turbulent velocity $\sim150-200$ km s$^{-1}$ is needed to produce a steepening frequency $\nu_s \sim 0.5-1$ GHz that is sufficient to explain the spectrum of the arcs observed between 1.4-3.0 GHz,   \citep[$\alpha \sim 2-3$,][]{Ignesti_2017}. However a decrease of a factor 2 in turbulent velocity would give a steepening frequency about 1 order of magnitude smaller. As a consequence of this, the steep spectrum emission in which the arcs are embedded should be brightest at lower radio frequencies. We note that also in this case explaining the observed shape of the arcs requires a specific 3-D geometry of the gas sloshing. The detailed investigation of the 3-D dynamics of the cold front would require tailored numerical simulations, which are beyond the aim of this paper. On the other hand, the presence of very steep spectrum emission between the arcs may be tested by deep LOFAR observations.

\section{Summary and Conclusion}

 In this paper we report on the analysis of new Chandra observations of the core region of the cluster A2626. The most important result of the paper is the discovery of a cold front. Interestingly this cold front is found in the proximity of the SW radio arcs suggesting a connection between the sloshing of the gas and the origin of the mysterious radio arcs in the cluster.

Radio arcs may be produced by the advection and gentle compression of seed ghost radio bubbles by the gas sloshing in the cold front or they may be the brightest regions of a mini-halo generated in the cold front region of the cluster. The latter hypothesis is somehow supported by numerical simulations of turbulent reacceleration in cold fronts. These simulations show filamentary radio structures arising near the surface of the cold fronts in the case that the reacceleration mechanisms are inefficient to generate emission at the observing frequencies. The expectation of this latter scenario is that diffuse inter-arcs emission with very steep spectrum should be brightest at lower frequencies.

In both cases the kite-like morphology of the radio arcs remain a puzzle and further steps require the use of tailored numerical simulations of sloshing cold fronts to address the 3-D geometry.

\section*{Acknowledgments}
%\vspace{-0.05in}
%We thank the referee for constructive comments that improved the
%presentation of the work. 
We thank the Referee for providing a constructive report that has improved the presentation of the results.
We thank P. Nulsen for providing the software required to produce the SB fit reported in Fig. 3 and M. Markevitch for useful discussion. 
AI thanks A. Botteon and G. Lanzuisi for useful discussions. MG, GB acknowledge partial support from PRIN-INAF
2014. EOS was supported in part by NASA grants GO5-16123X and GO6-17121X. CLS was supported in part by NASA Chandra grants GO4-15123X, GO5-16131X, and GO5-16146X. KWW was partially supported by Chandra grants GO5-16125X and GO6-17108X, and NASA ADAP grant NNH16CP10C. 

\vspace{-0.15in}
\bibliographystyle{aa}
\bibliography{bibliography}

\end{document}